\documentclass[11pt,epsf,letterpaper]{article}
\usepackage{setspace}
\usepackage{color}
\usepackage{amsmath}
\usepackage{amsfonts}
\usepackage{verbatim}
\usepackage{amssymb}
\usepackage{graphicx,bm}
\usepackage{graphicx,cite,enumerate}

\providecommand{\U}[1]{\protect\rule{.1in}{.1in}}
\newcommand{\be}{\begin{equation}}
\newcommand{\ee}{\end{equation}}
\newcommand{\beqn}{\begin{eqnarray}}
\newcommand{\eeqn}{\end{eqnarray}}
\newcommand{\pa}{\partial}

\begin{document}
\begin{flushright}
UAI-PHY-17/02
\end{flushright}
\vspace{0.8cm}
\begin{center}
\renewcommand{\thefootnote}{\fnsymbol{footnote}}
{\LARGE \bf Hairy AdS black holes with a toroidal horizon \\
\vskip 5mm
in 4D Einstein--nonlinear $\sigma $-model system}
\vskip 10mm

{\large {Marco Astorino$^{1}$\footnote{marco.astorino@gmail.com}, Fabrizio Canfora$^{2}$\footnote{canfora@cecs.cl}, Alex Giacomini$^{3}$\footnote{alexgiacomini@uach.cl},
Marcello Ortaggio$^{4,3}$\footnote{ortaggio@math.cas.cz}} \vspace{1cm} \\
%EndAName
{\small $^{1}$\textit{Universidad Adolfo Ibanez, Vi{\~n}a del Mar, Chile.}}\\
{\small $^{2}$\textit{Centro de Estudios Cient\'{\i}ficos (CECS), Casilla
1469, Valdivia, Chile.}}\\
{\small $^{3}$\textit{Instituto de Ciencias F\'{\i}sicas y Matem\'aticas,
Universidad Austral de Chile} }\\
{\small \textit{Edificio Emilio Pugin, cuarto piso, Campus Isla Teja,
Valdivia, Chile.}} \\
{\small $^{4}$\textit{Institute of Mathematics of the Czech Academy of
Sciences} }\\
{\small \textit{\v Zitn\' a 25, 115 67 Prague 1, Czech Republic.} }\\
}
\end{center}
\vskip 44mm
\begin{abstract}
An exact hairy asymptotically locally AdS black hole solution with a flat
horizon in the Einstein-nonlinear sigma model system in (3+1) dimensions is
constructed. The ansatz for the nonlinear $SU(2)$ field is regular
everywhere and depends explicitly on Killing coordinates, but in such a way
that its energy-momentum tensor is compatible with a metric with Killing
fields. The solution is characterized by a discrete parameter which has
neither topological nor Noether charge associated with it and therefore
represents a hair. A $U(1)$ gauge field interacting with Einstein gravity
can also be included. The thermodynamics is analyzed. Interestingly, the hairy black hole is always
thermodynamically favored with respect to the corresponding black hole with
vanishing Pionic field.
\end{abstract}

\newpage

\section{Introduction}

The nonlinear sigma model is a useful theoretical tool with applications
ranging from quantum field theory to statistical mechanics systems like
quantum magnetism, the quantum Hall effect, meson interactions, super fluid $%
^{3}$He, and string theory \cite{manton}. The most relevant application of
the $SU(2)$ non-linear sigma model in particle physics is the description of
the dynamics of Pions at low energy in 3+1 dimensions (see for instance \cite%
{example}; for a detailed review \cite{nair}). Consequently, the analysis of
the coupling of the nonlinear sigma model to General Relativity is extremely
important both from the theoretical and from the phenomenological point of
view. On the other hand, due to the complexity of the field equations
(which usually reduce to a non-linear system of coupled PDEs), the
Einstein-nonlinear sigma model system has been analyzed mostly relying on
numerical analyses (classic references are \cite{3,4,5,6,7}).

However, the recent generalization of the boson star ansatz to $SU(2)$%
-valued scalar fields (introduced in \cite%
{canfora,canfora2,canfora3,canfora4,yang1,canfora5,yang2,canfora6,canfora7,canfora8,canfora9}
and \cite{canfora10,canfora11,canfora12}) allows to construct also
non-trivial gravitating soliton solutions (see in particular \cite{canfora2}
and \cite{canfora6}) as well as to analyse explicitly thermodynamic
properties of solitons (see \cite{canfora12}). These are necessary
ingredients to build Pionic black holes with non-trivial thermodynamics.

Following the strategy devised in those references, we will construct a class
of analytic black hole solutions of the Einstein-$\Lambda $-nonlinear sigma
model system with intriguing geometrical properties. This
family of black holes possesses a flat horizon\footnote{%
Black holes with planar horizons have recently attracted a lot of attention
due to their applications in holography, see, e.g., \cite{AW,Gouteraux,CCPS}.%
} and a discrete hairy parameter. The thermodynamics can be
analysed explicitly. The first law is satisfied and an interesting feature
of these black holes is that the hairy solution has always less free energy
than the corresponding vacuum solution: thus, the present analysis suggests
that the coupling of Pions with the gravitational field can act as a sort of
catalysis for the Pions themselves. This family of black holes with flat
horizons can also be generalized to the case in which there is a $U(1)$
gauge field coupled to Einstein gravity. The interesting thermodynamical
features of these black holes remain in this case as well. This is qualitatively similar (in a phenomenological setting such as the
Einstein-Pions system) to the recent findings of \cite{herdeiro} in which a
family of asymptotically flat black holes with non-Abelian hair has been
presented which are thermodynamically favoured over the Reissner-Nordstr\"om solution.

The paper is organized as follows. In section~\ref{sec_action}, the
Einstein-Nonlinear $\sigma $-model is introduced and a convenient
parametrization is described. The exact hairy black hole solutions are
constructed in section \ref{sec_solutions}. In section~\ref{sec_thermod}, the
thermodynamic behavior of the solutions is analyzed. In the last section~\ref{sec_concl}, some conclusions are drawn.\\

\section{The action}

\label{sec_action}

We consider the Einstein-non-linear sigma model system in four dimensions,
with a possible cosmological constant. This describes the low-energy
dynamics of pions, whose degrees of freedom are encoded in an $SU(2)$
group-valued scalar field $U$ \cite{nair}. The action of the system is
\begin{equation}
S=S_{\mathrm{G}}+S_{\mathrm{Pions}},  \label{actiontotal}
\end{equation}%
where the gravitational action $S_{\mathrm{G}}$ and the nonlinear sigma
model action $S_{\mathrm{Pions}}$ are given by
\begin{align}
S_{\mathrm{G}}=& \frac{1}{16\pi G}\int d^{4}x\sqrt{-g}(\mathcal{R}-2\Lambda
),  \label{sky1} \\
S_{\mathrm{Pions}}=& \frac{K}{4}\int d^{4}x\sqrt{-g}\mathrm{Tr}\left( R^{\mu
}R_{\mu }\right) .  \label{sky2}
\end{align}%
Here we have defined
\begin{equation}
R_{\mu }=U^{-1}\nabla _{\mu }U\ , \qquad U\in SU(2)\ ,
\end{equation}
while $\mathcal{R}$ is the Ricci scalar, $G$ is Newton's constant, $\Lambda$
is the cosmological constant and the parameter $K$ is positive. In our conventions $c=\hbar =1$, the space-time signature is $%
(-,+,+,+)$ and Greek indices run over space-time. In the case of the non-linear sigma model on flat space-times, \
the coupling constant $K$ is fixed experimentally. On the other hand, its true
meaning in the context of AdS physics is to introduce a new length scale.
Indeed, it is natural to expect that the ratio of this new length scale with
the AdS radius will be a relevant quantity in the thermodynamics of the
black holes which will be analyzed in the following sections. Our results
confirm this expectation (we thank the anonymous referee for this comment).

The resulting Einstein equations are
\begin{equation}
G_{\mu \nu }+\Lambda g_{\mu \nu }=8\pi GT_{\mu \nu },  \label{einstein}
\end{equation}%
where $G_{\mu \nu }$ is the Einstein tensor and the energy-momentum tensor
is
\begin{equation}
T_{\mu \nu }=-\frac{K}{2}\mathrm{Tr}\left( R_{\mu }R_{\nu }-\frac{1}{2}%
g_{\mu \nu }R^{\alpha }R_{\alpha }\right) \,.  \label{timunu1}
\end{equation}%
For the nonlinear $SU(2)$ sigma model, $T_{\mu \nu }$ can be seen to satisfy
the dominant and strong energy conditions \cite{gibbons2003}. Finally, the
matter field equations are
\begin{equation}
\nabla ^{\mu }R_{\mu }=0.  \label{nonlinearsigma1}
\end{equation}

In the following, it will be useful to write $R_{\mu }$ as
\begin{equation}
R_{\mu }=iR_{\mu }^{j}\sigma_{j} ,  \label{Ri}
\end{equation}
where $\sigma _{j}$ are the Pauli matrices. Furthermore, we adopt the
standard parametrization of the $SU(2)$-valued scalar $U(x^{\mu })$
\begin{equation}
U^{\pm 1}(x^{\mu })=Y^{0}(x^{\mu })\mathbb{\mathbf{I}}\pm iY^{j}(x^{\mu
})\sigma_{j}\ , \qquad \left( Y^{0}\right) ^{2}+Y^{i}Y_{i}=1\,,
\label{standnorm}
\end{equation}
where $\mathbb{\mathbf{I}}$ is the $2\times 2$ identity matrix. From \eqref{Ri} one
thus finds
\begin{equation}
R_{\mu }^{i}=\varepsilon ^{ijk}Y_{j}\nabla _{\mu }Y_{k}+Y^{0}\nabla
_{\mu}Y^{i}-Y^{i}\nabla _{\mu }Y^{0} .  \label{Ri_2}
\end{equation}

Defining the quadratic combination
\begin{equation}
\mathcal{S}_{\mu \nu }:=\delta _{ij}R_{\mu }^{i}R_{\nu }^{j}=\nabla_{\mu
}Y^{0}\nabla _{\nu }Y^{0}+\nabla_{\mu }Y^{i}\nabla _{\nu
}Y^{i}=G_{ij}(Y)\nabla _{\mu }Y^{i}\nabla _{\nu }Y^{j}\ ,  \label{cuadra1}
\end{equation}%
with
\begin{equation}
G_{ij}:=\delta _{ij}+\frac{Y_{i}Y_{j}}{1-Y^{k}Y_{k}}\   \label{intmetric}
\end{equation}%
being the $S^3$ metric of the target space,\footnote{%
For the special field configuration $Y^0=0$, eq.~\eqref{intmetric} should
be replaced by $G_{ij}:=\delta _{ij}$. This case will not be considered in
this paper.} we obtain that the action (\ref{sky2}) reads
\begin{equation}
S_{\mathrm{Pions}}=-K\int d^{4}x\sqrt{-g}\left[ \frac{1}{2}G_{ij}(\nabla
_{\mu }Y^{i})(\nabla ^{\mu }Y^{j})\right] ,
	\label{action2}
\end{equation}
while the energy-momentum tensor (\ref{timunu1}) takes the form
\begin{equation}
T_{\mu \nu }=K\left( \mathcal{S}_{\mu \nu }-\frac{1}{2}g_{\mu \nu }\mathcal{S%
}\right) \ .   \label{timunu2}
\end{equation}

The second of \eqref{standnorm} means that $Y^{I}:=(Y^{0},Y^{i})$ define a
round unit 3-sphere in the internal space. A useful set of coordinates $%
(H,A,G)$ in the internal space is defined by
\begin{eqnarray}
&&Y^{0}=\cos H\sin A,\qquad Y^{1}=\sin H\cos G,  \notag \\
&&Y^{3}=\cos H\cos A,\qquad Y^{2}=\sin H\sin G,  \label{HAG_def}
\end{eqnarray}%
where $H\in \lbrack 0,\pi /2]$, while $A\in \left[ 0,2\pi k_{1}\right] $, $%
G\in \left[ 0,2\pi k_{2}\right] $ are both periodic Killing coordinates of $%
S^{3}$, with $k_{1}$ and $k_{2}$ positive integers (there is clearly some redundancy in this choice of the periodicities -- the standard choice $k_1=k_2=1$ already covers the whole $S^3$ -- but this will become physically meaningful later on).  In \cite{canfora6}\ and
\cite{canfora12}\ it has been shown that similar parametrizations are extremely
useful both in curved and in flat spaces. The tensor $\mathcal{S}_{\mu \nu }$\ \eqref{cuadra1}
takes the form
\begin{equation}
\mathcal{S}_{\mu \nu }=(\nabla _{\mu }H)(\nabla _{\nu }H)+\cos ^{2}H(\nabla
_{\mu }A)(\nabla _{\nu }A)+\sin ^{2}H(\nabla _{\mu }G)(\nabla _{\nu }G).
\label{S_H}
\end{equation}

If one defines the following combinations of the Killing coordinates
\begin{equation}
\Phi _{+}=G+A,\qquad \Phi _{-}=G-A,
\end{equation}%
the field equations~(\ref{nonlinearsigma1}) can be written in the compact
form\footnote{%
It is useful to recall the simple identity $\left( g^{\mu \nu }hf_{,\nu
}\right) _{;\mu }=\nabla h\cdot \nabla f+h\square f$.}
\begin{eqnarray}
&&\square H-\frac{1}{2}\sin (2H)\nabla \Phi _{+}\cdot \nabla \Phi _{-}=0,
\label{D2H} \\
&&\sin (2H)\square \Phi _{+}+2\nabla H\cdot \left[ \nabla \Phi _{-}+\cos
(2H)\nabla \Phi _{+}\right] =0,  \label{D2ga} \\
&&\sin (2H)\square \Phi _{-}+2\nabla H\cdot \left[ \nabla \Phi _{+}+\cos
(2H)\nabla \Phi _{-}\right] =0.  \label{D2ph}
\end{eqnarray}

$SU(2)$-valued scalar fields may possess a non-trivial topological charge
which, mathematically, is a suitable homotopy class or winding number $W$.
Its explicit expression as an integral over a suitable three-dimensional
hypersurface $\Sigma $ can be found, e.g., in \cite{manton}. However, in the present
paper we will consider only configurations with $W=0$.

\section{Toroidal black hole solutions}

\label{sec_solutions}

\subsection{Uncharged black hole}

\label{subsec_bh}

We now analyze configuration which are a natural non-topological generalization of the ansatz of
\cite{canfora6,canfora12}.

We will consider a static spacetime with a flat base manifold and the
following diagonal metric
\begin{equation}
ds^{2}=-f\left( r\right) dt^{2}+h\left( r\right) dr^{2}+r^{2}\left( d\theta
^{2}+d\varphi ^{2}\right) \ .  \label{flat3}
\end{equation}%
In this geometry, for the matter field~\eqref{HAG_def} we choose the
\textquotedblleft adapted\textquotedblright\ configuration
\begin{equation}
H=\frac{\pi }{4},\qquad A=\theta ,\qquad G=\varphi ,  \label{flat1}
\end{equation}%
which is consistent with \eqref{HAG_def} if we set the periodicity of the
angular spacetime coordinates as
\begin{equation}
\theta\in[0,2\pi k_1] , \qquad \varphi\in[0,2\pi k_2] ,
\end{equation}
making the base manifold in \eqref{flat3} a (flat) 2-torus.\footnote{%
Here the integers $k_1$ and $k_2$ are used
to define the identifications of points in the physical spacetime and are
therefore not redundant anymore (as opposed to \eqref{HAG_def}). Also note
that the cylindrical [and planar] topology is included in the formal limit $%
k_1\to+\infty$ [and $k_2\to+\infty$].} It is easy to see that, thanks to %
\eqref{flat3} and \eqref{flat1}, the field equations~\eqref{D2H}--%
\eqref{D2ph} are satisfied identically. In addition, by construction we have
$W=0$ (since the field \eqref{flat1} is 2-dimensional), as we required. We
further note that this Pionic configuration has also zero Noether charge
associated with the global isospin symmetry of the model. Indeed, the
Noether charge is proportional to the spatial integral of the time-component
of the Nother current associated with the global Isospin symmetry of the
model. In the present case, due to the fact that the $SU(2)$-valued field
does not depend on time, its Noether charge vanishes identically.

One still needs to solve Einstein's equations~\eqref{einstein} with %
\eqref{timunu2}, which here becomes
\begin{equation}
T_{\mu\nu}=\frac{K}{2r^2}(f\delta^t_\mu\delta^t_\nu-h\delta^r_\mu\delta^r_%
\nu) .  \label{Tmunu_reduced}
\end{equation}
Eq.~\eqref{einstein} is thus solved by \eqref{flat3} (up to a constant
rescaling of $t$) with
\begin{equation}
\frac{1}{h(r)}=f(r)=-4\pi G K-\frac{\mu}{r}-\frac{\Lambda}{3}r^2 ,
\label{BH_flat}
\end{equation}
where $\mu$ is an integration constant.

Geometric features of the spacetime are described more transparently if we
introduce two \textquotedblleft normalized\textquotedblright\ coordinates $%
x,y\in \lbrack 0,1]$ such that
\begin{equation}
\theta =2\pi k_{1}x,\qquad \varphi =2\pi k_{2}y\ ,  \label{xy}
\end{equation}%
and additionally rescale
\begin{equation}
t\mapsto 2\pi t,\qquad r\mapsto \frac{r}{2\pi},\qquad \mu
\mapsto \frac{\mu }{(2\pi)^{3}}\ .
\end{equation}%

Defining the convenient parameters\footnote{%
Here we are interested in solutions in AdS, but in \eqref{BH_flat} $\Lambda $
can take any sign or vanish.}
\begin{equation}
	b^{2}=16\pi^3GK \ ,\qquad \ell^2= -\frac{3}{\Lambda} \ ,
\end{equation}%
the final form of the solution is given by the field \eqref{flat1}, %
\eqref{xy} in the spacetime
\begin{eqnarray}
 & & ds^{2}=-f\left( r\right) dt^{2}+\frac{dr^{2}}{f\left( r\right) }%
+r^{2}\left(k_1^{2}dx^{2}+k_2^{2}dy^{2}\right) \ ,  \label{final-metric} \\
&&f(r)=-b^2-\frac{\mu}{r}+\frac{r^2}{\ell^2} \ .  \label{f_pions}
\end{eqnarray}%
Note that $b^2$ is not an integration constant, but is fixed by the coupling constants of the theory.

Let us mention a possible alternative parametrization of this solution. Rescaling
\be
  k_1=\epsilon \tilde k_1 , \qquad k_2=\epsilon \tilde k_2 , \qquad \mu=\tilde \mu/\epsilon^3 , \qquad r=\tilde r/\epsilon , \qquad t=\epsilon \tilde t ,
\ee
one obtains
\begin{eqnarray}
 & & ds^{2}=-f\left( \tilde r\right) d\tilde t^{2}+\frac{d\tilde r^{2}}{f\left( \tilde r\right) }%
+{\tilde r}^{2}\left(\tilde k_1^{2}dx^{2}+\tilde k_2^{2}dy^{2}\right) \ ,  \label{final-metric_2} \\
&&f(\tilde r)=-\epsilon^2 b^2-\frac{\tilde \mu}{\tilde r}+\frac{\tilde r^2}{\ell^2} \ .  \label{f_pions_2}
\end{eqnarray}%
This allows one to set $\epsilon=0$, if desired, thus recovering the known toroidal vacuum solution \cite{Lemos95,HuaLia95,Mann97,BLP,Vanzo} -- this limit is however only ``formal'', since the Pionic field \eqref{flat1}, \eqref{xy} degenerates discontinuously to a single point in the internal space.

From now on we stick to the parametrization \eqref{final-metric}, \eqref{f_pions}. The base space is a flat 2-torus with Teichm\"{u}ller parameter $ik_1/k_2$ (cf., e.g., \cite{Vanzo}). Metric~\eqref{final-metric} is similar to the well-known vacuum topological
black holes, for which the base manifold can be spherical, hyperbolic or
toroidal \cite{Lemos95,HuaLia95,Mann97,BLP,Vanzo}. However,
while in the absence of Pions the constant term in the lapse function is
fixed by the curvature of the base manifold (i.e., $+1,-1,0$ in the
spherical, hyperbolic and toroidal case, respectively), in \eqref{f_pions}
it takes a negative value despite the fact that the base is flat. As a consequence, the base space area
\be
 \sigma=k_1k_2 ,
\ee
plays the role of an extra (discrete) parameter that cannot be rescaled away by a redefinition of $\mu$ (as opposed to toroidal black holes in vacuum \cite{Lemos95,HuaLia95,Mann97,BLP,Vanzo}). Recalling that the present Pionic configuration has neither topological
nor Noether charge, $\sigma$ can be considered as an integer Pionic hair of
the black hole.

Because of the form of the lapse function~\eqref{f_pions}, one can easily adapt to the present context
the discussion of the causal structure and horizons of hyperbolic vacuum black holes given in \cite{BLP,Vanzo}
(to which we refer for more details). First, since $r=0$ is a curvature
singularity, we restrict ourselves to the range $r>0$. The discriminant
of $f(r)=0$ reads (up to an overall positive factor) $\Delta
=4\ell^2b^{6}-27\mu ^{2}$, while the three roots satisfy $r_{1}r_{2}r_{3}=\mu
\ell^2>0$ and $r_{1}+r_{2}+r_{3}=0$. It follows that there cannot be three
real roots with the same sign, and that there are no real positive roots
(only) when $\Delta $ and $\mu $ are both negative. To be more precise, let
us define the (positive) critical value of $\mu $ (such that $\Delta =0$ for
$\mu =\pm \mu _{c}$, cf. \cite{BLP,Vanzo})
\begin{equation}
\mu _{c}=\frac{2\sqrt{3}}{9}\ell b^{3}.
\end{equation}%
There exists a unique, positive simple root $r_{+}$ for $\mu \geq 0$, two
distinct positive roots $r_{+}$ and $r_{-}$ (with $r_{+}>r_{-}$) for $-\mu
_{c}<\mu <0$, a double positive root for $\mu =-\mu _{c}$, and no positive
roots for $\mu <-\mu _{c}$, the latter case thus describing a naked
singularity. The critical value $\mu =-\mu _{c}$ gives rise to a degenerate
Killing horizon at $r=r_{e}\equiv \ell b/\sqrt{3}$, however this spacetime does
not describe a black hole \cite{BLP,Vanzo}. In order to have a black hole
solution, one thus needs to take $\mu >-\mu _{c}$. Therefore $r_+$ is a monotonically increasing function of $\mu$ and $r_+> r_e$.

\subsection{Charged black hole}

The generalization of the above model to the presence of a $U(1)$ gauge
field minimally coupled to General Relativity is direct. The main motivation
for this extension consists in analyzing whether or not the thermodynamic
behaviour disclosed in the Einstein-nonlinear sigma model system (discussed
in section~\ref{sec_thermod}) resists after the inclusion of further reasonable
matter fields.

More precisely, we add to the action (\ref{actiontotal}) the electromagnetic
term
\begin{equation}
S_{\mathrm{Max}}=-\frac{1}{16\pi G}\int d^{4}x\sqrt{-g}\left( F_{\mu \nu
}F^{\mu \nu }\right) \ ,
\end{equation}%
where $F_{\mu \nu }=\partial _{\mu }A_{\nu }-\partial _{\nu }A_{\mu }$,
which adds to the energy-momentum tensor \eqref{timunu2} the usual Maxwell
term
\begin{equation}
\frac{1}{4\pi G}\left( F_{\mu \rho }F_{\nu }^{\ \rho }-\frac{1}{4}g_{\mu \nu }F_{\rho
\sigma }F^{\rho \sigma }\right) \ .
\end{equation}%
We will focus on static electric fields defined by the potential
\begin{equation}
A_{\mu }=\left( -\frac{q}{r},0,0,0\right) \ .  \label{A}
\end{equation}%
This form of $A_{\mu }$ has the property of being compatible with the metric
(\ref{final-metric}), i.e., the Maxwell equations
\begin{equation*}
\partial _{\mu }(\sqrt{-g}F^{\mu \nu })=0\
\end{equation*}%
are automatically satisfied. Solving Einstein's equations with the ansatz %
\eqref{flat3}, \eqref{flat1} and \eqref{A} shows that the contribution of
the electric field consists just in an upgrading of the the lapse function~%
\eqref{f_pions} to
\begin{equation}
f(r)=-b^{2}-\frac{\mu }{r}+\frac{r^{2}}{\ell^2}+\frac{q^{2}}{{r}^{2}}\ .
\label{f_charged}
\end{equation}%
Of course, the presence of an electric field modifies the structure of the
black hole horizons. In general in this charged case we have both an inner horizon $r_-$ and an outer horizon $r_+$, see \cite{BLP} for a related discussion.

\section{Thermodynamic behavior of the asymptotically locally AdS black hole
solution}

\label{sec_thermod}

\subsection{Uncharged black hole}

Let us begin by analyzing the electrically uncharged case. We employ the Euclidean approach \cite{GibHaw77} and thus replace $t\mapsto it$ in \eqref{final-metric}. The temperature is given by the inverse of the Euclidean time period, i.e.,
\begin{equation}
T\equiv \beta^{-1} = \frac{1}{4\pi }f^{\prime }(r_{+})=\frac{1}{4\pi }\left( \frac{\mu }{%
r_{+}^{2}}+\frac{2r_{+}}{\ell^2}\right) \ ,
\end{equation}
or, expressing $\mu=r_+(r_+^2/\ell^2-b^2)$ as a function of $r_{+}$, by
\begin{equation}
T=\frac{1}{4\pi }\left( \frac{-b^{2}}{r_{+}}+\frac{3r_{+}}{\ell^2}\right) \ .
\label{T_2}
\end{equation}%
This is a monotonically increasing function of $r_+$ (and thus of $\mu$), with $T\in[0,+\infty)$ vanishing for the extremal solution with $\mu=-\mu_c$. Inverting this relation admits only one positive root for $r_+$, which expresses the horizon radius as a function of $T$
\begin{equation}
r_+ =\frac{\ell^2}{3}\left( 2\pi T +\sqrt{4\pi^2T^2+\frac{3b^2}{\ell^2}}\right) ,
 \label{r+}
\end{equation}
therefore no bifurcations of the free energy similar to the Hawking-Page \cite{HawPag83} phase transition are possible here.

The Euclidean action (which we denote by $I$ to distinguish it from its Lorentzian counterpart) is a sum of three contributions, namely
\be
 I=I_{bulk}+I_{surf}+I_{ct} ,
 \label{action_eucl}
\ee
with (recall \eqref{cuadra1}--\eqref{action2})
\beqn
 & & I_{bulk}=-\frac{1}{16\pi G}\int_{\cal M} d^{4}x\sqrt{g}\left(\mathcal{R}+\frac{6}{\ell^2}\right)+\frac{K}{2}\int_{\cal M} d^{4}x\sqrt{g}G_{ij}g^{\mu\nu}\pa
_\mu Y^{i}\pa_\nu Y^{j} , \\
 & & I_{surf}=-\frac{1}{8\pi G}\int_{\pa\cal M}d^{3}x\sqrt{h}K_B ,
\eeqn
where $h_{ab}$ is the metric induced on the boundary $\pa\cal M$ at $r=r_B$, and $K_B$ is the trace of the extrinsic curvature of $\pa\cal M$ as embedded in $\cal M$. For $r_B\to+\infty$, both $I_{bulk}$ and $I_{surf}$ are divergent when evaluated on our solutions, nevertheless \eqref{action_eucl} can be made finite by an appropriate definition of the counterterm action $I_{ct}$. The latter consists of boundary integrals involving scalars constructed only from boundary quantities. At this purpose, taking inspiration from \cite{CCPS}, where similar matter field was studied, we naturally generalise the counterterm action to the more general collection of scalar fields, we are considering, in the following way
\be \label{Icounterterm}
  I_{ct}=\frac{1}{8\pi G}\int_{\pa\cal M}d^{3}x\sqrt{h}\frac{\ell}{2}\left(\mathcal{R}_B+\frac{4}{\ell^2}\right) -\frac{K}{2}\int_{\cal\pa M} d^{3}x\sqrt{h}\,\ell G_{ij}h^{ab}\pa_a Y^{i}\pa_b Y^{j} ,
\ee
where $\mathcal{R}_B$ is the Ricci scalar of the boundary metric. When the sigma model coupling of the scalar fields is trivialised to become a kinetic interaction, i.e.  $G_{ij}=\delta_{ij}$, the axion contertem of \cite{CCPS} is, indeed, recovered from (\ref{Icounterterm}). The AdS gravitational part of the above counterms is well-known \cite{BalKra99} (see also \cite{EmpJohMye99,Mann99}). Taking the limit $r_B\to+\infty$ gives the renormalized action
\be
 I=-\frac{\beta\sigma r_+}{16\pi G\ell^2}\left(r_+^2+b^2\ell^2\right) .
\ee
It is easy to see that $I\sim T^2$ for a large $T$ (similarly as for hyperbolic vacuum  black holes \cite{Vanzo}).

The energy $E=-\pa_\beta\log Z$, the entropy $S=(1-\beta\pa_\beta)\log Z$ and the free energy $F=-T\log Z$ of the solutions in the semiclassical approximation $\log Z\approx -I$ thus read
\be
 E=\frac{\sigma r_+}{8\pi G}\left(\frac{r_+^2}{\ell^2}-b^2\right)=\frac{\sigma\mu}{8\pi G} , \qquad S=\frac{\sigma r_+^2}{4G} , \qquad F=-\frac{\sigma r_+}{16\pi G\ell^2}\left(r_+^2+b^2\ell^2\right) .
\label{ESF}
\ee

The area law is thus clearly obeyed. By defining the the mass as
 \be
	M=E ,
	\label{mass}
\ee
it is easy to check with~\eqref{T_2} that the first law of black hole thermodynamics is satisfied, i.e.,
\begin{equation}
T\delta S=\frac{\sigma}{8\pi G}\left(-b^{2}+\frac{3r_{+}^{2}}{\ell^2}\right)\delta
r_{+}=\delta M\ .
\end{equation}

The local thermodynamic stability with respect to thermal fluctuations is given by
the positivity of the heat capacity function $C$, i.e.,
\begin{equation}  \label{spech-heat}
C := T \left( \frac{\partial S}{\partial T} \right) = T \left( \frac{%
\partial S}{\partial r_+} \right) \left( \frac{\partial T}{\partial r_+}
\right)^{-1} = \ \frac{r^2_+ \tau (3 r_+^2 - b^2 \ell^2)}{2 G(b^2\ell^2+3r_+^2)} \ .
\end{equation}
Since black hole solutions satisfy $r_+> \frac{b \ell}{\sqrt{3}}$ (section~\ref{subsec_bh}), they are stable under temperature fluctuation (this is also true
in the absence of the Pionic matter, cf. \cite{BLP,Vanzo}). In other words, $M$ is a monotonically increasing function of $T$ (see also \eqref{M(T)} below).

We further note that \eqref{mass} gives a partly negative mass spectrum (similarly as obtained in \cite{EmpJohMye99} for hyperbolic vacuum black holes -- cf. \cite{BLP,Vanzo} for a different approach). Note also that $M$ and $S
$ depend not only on the radial length $r_+$ (i.e., on the integration constant $\mu$), but also on the angular periods via $\sigma$.
It can thus be seen that, for a given $M$, larger entropies are attained for larger (discrete) values of $\sigma$. In addition, for a given $T$ (i.e., for a fixed $r_+$) there exists a discrete infinity of black holes parametrized by $\sigma$, for which the functions \eqref{ESF} take different values, namely

\beqn
  & & M =\frac{\ell \sigma}{108\pi G}\left( 2\pi T \ell + \sqrt{4\pi^2 T^2 \ell^2 +3b^2}%
\right)\left(4\pi^2 T^2 \ell^2 + 2\pi T \ell \sqrt{4\pi^2 T^2 \ell^2 +3b^2}-3b^2
\right) , \label{M(T)}  \\
	& & S = \frac{\ell^4\sigma}{9 G}\left(2\pi^2 T^2 +\pi T \sqrt{4\pi^2 T^2 +\frac{3b^2}{%
\ell^2}}+\frac{3b^2}{4\ell^2}\right) , \label{S(T)} \\
	& & F = -\frac{\ell \sigma}{108\pi G}\left( 2\pi T \ell + \sqrt{4\pi^2 T^2 \ell^2 +3b^2}%
\right)\left(3b^2 +2\pi^2 T^2 \ell^2 + \pi T \ell \sqrt{4\pi^2 T^2 \ell^2 +3b^2}
\right) . \label{F(T)}
\eeqn

At high temperatures, $M\sim \sigma T^3$, $S\sim \sigma T^2$ and $F\sim -\sigma T^3$.

It is interesting to compare this black hole solution dressed with a matter
field with the corresponding vacuum solution (i.e., with $b=0$). The entropy
of the toroidal vacuum black hole is \cite{BLP,Vanzo}
\begin{equation}
S_0 = \frac{l^4\sigma}{9 G} 4\pi^2 T^2 \ .
\end{equation}
Clearly $S_0<S$, i.e., at equal temperature the vacuum black hole has
always lower entropy than the dressed solution. In order to see which black
hole is thermodynamically favored at a given temperature, it is necessary to
compare the respective free energies. From \eqref{F(T)}, it is obvious that the free energy is more negative for
$b^2\neq0$, i.e., the solution with Pionic matter fields is
thermodynamically favored at any temperature.

This result is remarkable, since usually black holes dressed with matter
fields in asymptotically locally AdS spacetimes are thermodynamically
favored over their matter-free counterparts only for low temperatures (see, e.g., \cite{Gubser08,MarTroZan04}). Recently, results qualitatively similar to ours have been found
in \cite{herdeiro}. However, this reference obtained numerically solutions of a higher order version of Yang-Mills theory minimally
coupled to General Relativity which are asymptotically flat -- here we have constructed analytic results for Pions minimally coupled with AdS gravity.

\subsection{Charged black hole}

The extension to the electrically charged case is straightforward. Now the
lapse function is given by \eqref{f_charged}. The electric charge reads $Q =
\frac{q\sigma}{4\pi}$ and the mass is the same as in the uncharged case.
Expressing $M= M(r_+, Q)$ as a function of the horizon radius and of the
electric charge, one readily finds that the first law of thermodynamics
\begin{equation}
\delta M = T\delta S + \Phi \delta Q
\end{equation}
is again fulfilled. The Coulomb electric potential $\Phi$ at the horizon is
defined, as usual, as
\begin{equation} \label{Phi}
\Phi = \chi^\mu A_\mu \Big|_{r_\infty} - \chi^\mu A_\mu \Big|_{r_+} = \frac{q%
}{r_+} \quad ,
\end{equation}
where $\chi^\mu$ represents the Killing vector $\partial_t$, while the temperature of the charged solution is
\begin{equation} \label{TrpQ}
T = \frac{1}{4\pi }\left( \frac{\mu }{%
r_{+}^{2}}+\frac{2r_{+}}{\ell^2} - \frac{q^2}{r_+^3}\right) \quad = \frac{1}{4 \pi r_+^3} \left( \frac{3 r_+^4}{\ell^2} - b^2 r_+^2 -q^2 \right) .
\end{equation}
In order to check the thermodynamic stability of this solution, we
compare the free energy of the electrically charged black hole with Pionic
field with the electrically charged black hole without Pionic field.
The free energy in the gran canonical ensamble, i.e. when both the temperature $T$ and the Coulomb potential $\Phi$ are considered fixed, for the charged pionic solution is given by the Gibbs free energy
\begin{equation}
                  F =   M - T S - \Phi Q   \quad .
\end{equation}

When $\mu$ is expressed in terms of $r_+$, the free energy becomes
\begin{equation}
                    F(r_+,q) = - \frac{r_+^4 + \ell^2(q^2+b^2r_+^2)}{16 \pi r_+ \ell^2} \sigma \quad .
\end{equation}
In the electrically charged case, it is convinient to write the free energy as a
function of the intensive parameters such as the temperature $T$ and the Coulomb potential $\Phi$. This can be done inverting eqs. (\ref{Phi}) and (\ref{TrpQ}). Again, there is only one positive root of the radial position of the event horizon as a function of the temperature $r_+(T)$
\begin{equation}
                           r_+(T) = \frac{2}{3} \pi \ell^2 T + \frac{\ell}{3} \sqrt{4 \pi \ell^2 T^2 + 3 (b^2 +\Phi^2)}   .
\end{equation}
Finally the Gibbs free energy reads as follows
\begin{equation} \label{freeQ}
                        F(T,\Phi) =  \frac{-\sigma \ell}{108\pi} \left(2\pi \ell T+\sqrt{4\pi^2 \ell^2 T^2 + 3 b^2 + 3\Phi^2} \right) \left[ 3 b^2  +\pi \ell T \left( 2\pi \ell T+\sqrt{4\pi^2 \ell^2 T^2 + 3 b^2 + 3\Phi^2} \right) \right] .
\end{equation}
This expression for the free energy is the generalisation of the uncharged one, given in eq. (\ref{F(T)}), which can be straightly recovered in the vanishing electric potential limit.

\begin{figure}[h!]
  \centering
 \includegraphics[scale=1]{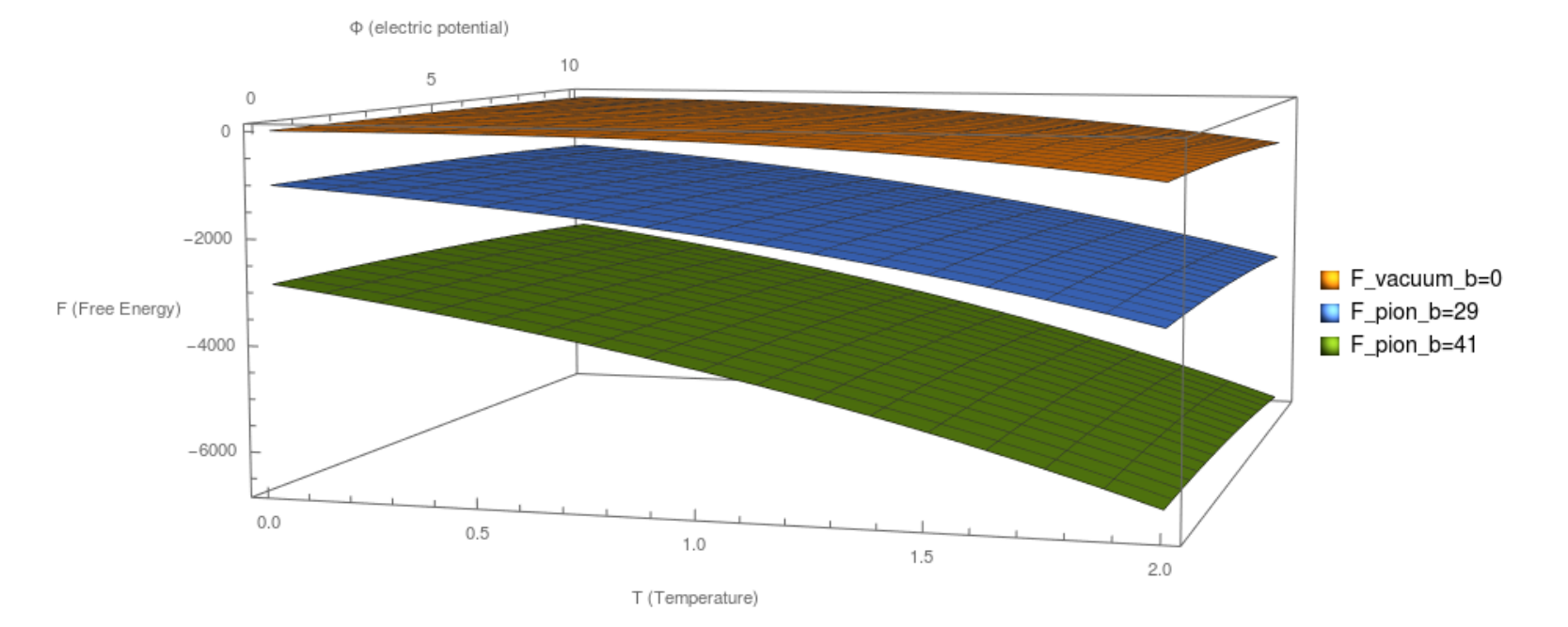}
  \caption{\small Gibbs free energy $F(T,\Phi)$ comparison between the electrovacuum (yellow) and the pionic (blue and green) black holes as function of the temperature $T$  and electric potential $\Phi$ for some fixed numerical values of the cosmological constant and the base manifold area $\ell=\pi,\sigma=1$. The picture does not change qualitatively for different values of the chosen pionic parameter $b$ (chosen as 29 and 41 respectively in blue and the green surfaces): the hairy configuration is always preferred at a given temperature and electric potential since its free energy is lower.}
\label{free-energy}
 \end{figure}

Again, the free energy of the dressed black hole turns out to be more more negative at the same temperature and electric potential, than the one of the black hole without Pionic matter field, as can be seen from eq.~(\ref{freeQ}) and in Figure~\ref{free-energy}. Therefore the charged pionic solution is always preferred. \\

\vspace{1cm}

\section{Conclusions}

\label{sec_concl}

We constructed exact static solutions of the four-dimensional Einstein
theory, minimally coupled to the nonlinear sigma model. A $U(1)$ gauge field
interacting with Einstein gravity can also be included. The ansatz for the
nonlinear sigma model depends explicitly on the Killing coordinates, but in
such a way that its energy-momentum tensor is compatible with the Killing
fields of the metric. The most interesting result is the construction of an
analytic hairy black hole sourced by a clouds of Pions in the Einstein
non-linear sigma model system with a negative cosmological constant. Such an
asymptotically locally AdS black hole has a flat horizon. The hairy
parameter is discrete due to the $SU(2)$ structure of the matter field --
this is thus an explicit example of a discrete primary hair. The
thermodynamic analysis reveals that the hairy black hole is always favoured
with respect to the black hole without the Pions cloud. The inclusion of a $U(1)$ gauge field minimally
coupled to General Relativity does not break this picture. This is a highly
phenomenological realization in the Einstein-Pions system of a behaviour similar to the recent
findings of \cite{herdeiro} in a different context.

\subsection*{Acknowledgements}
{\small
M.O. is grateful to Roberto Emparan for a useful discussion. This work has been partially funded by the FONDECYT grants 1160137 (F.C.), 1150246 (A.G.), 11160945 (M.A.), the CONICYT grant DPI20140053 (A.G.) and research plan RVO: 67985840 and research grant GA\v CR 14-37086G (M.O.). M.A. is supported by Conicyt - PAI grant n$^o$ 79150061. The stay of M.O. at Instituto de Ciencias F\'{\i}sicas y Matem\'aticas, Universidad Austral de Chile has been supported by CONICYT PAI ATRACCI{\'O}N DE CAPITAL HUMANO AVANZADO DEL EXTRANJERO Folio 80150028. The Centro de Estudios Cient\'{\i}ficos (CECs) is
funded by the Chilean Government through the Centers of Excellence Base
Financing Program of Conicyt.}\\

%\bibliographystyle{unsrt}

%\bibliography{bibl}

\begin{thebibliography}{99}
\bibitem{manton} N.~Manton and P.~Sutcliffe, \textit{Topological Solitons},
Cambridge University Press, Cambridge, 2007.

\bibitem{example} J. Gasser, H. Leutwyler, \textit{Nucl. Phys.}\textbf{\ B
250}, 465 (1985).

\bibitem{nair} V. P. Nair, ``\textit{Quantum Field Theory: A Modern
Perspective}'', Springer (2005).

\bibitem{derrick} G.H. Derrick, \textit{J. Math. Phys}. \textbf{5}:
1252--1254 (1964).

\bibitem{bosonstar} D. J. Kaup, \textit{Phys. Rev}.\textbf{172}, 1331--1342
(1968).

\bibitem{bosonstar2} S. L. Liebling, C. Palenzuela, \textit{Living Rev.
Relativity} \textbf{15}, (2012), 6.

\bibitem{3} S. Droz, M. Heusler, and N. Straumann, Phys. Lett. B 268, 371
(1991).

\bibitem{4} H. Luckock and I. Moss, Phys. Lett. B 176, 341 (1986).

\bibitem{5} S. Droz, M. Heusler, and N. Straumann, Phys. Lett. B 271, 61
(1991).

\bibitem{6} N. K. Glendenning, T. Kodama, and F. R. Klinkhamer, Phys. Rev. D
38, 3226 (1988); B.M.A.G. Piette and G. I. Probert, Phys. Rev. D 75, 125023
(2007); G.W. Gibbons, C. M. Warnick, and W.W. Wong, J. Math. Phys. (N.Y.)
52, 012905 (2011); S. Nelmes and B. M. A. G. Piette, Phys. Rev. D 84, 085017
(2011).

\bibitem{7} P. Bizon and T. Chmaj, Phys. Rev. D 58, 041501 (1998); P. Bizon,
T. Chmaj, and A. Rostworowski, Phys. Rev. D 75, 121702 (2007); S. Zajac,
Acta Phys. Pol. B 40, 1617 (2009); 42, 249 (2011).

\bibitem{canfora} F. Canfora, P.~Salgado-Rebolledo, \textit{Phys. Rev.}
\textbf{D 87}, 045023 (2013).

\bibitem{canfora2} F. Canfora, H.~Maeda, \textit{Phys. Rev.} \textbf{D 87},
084049 (2013).

\bibitem{canfora3} F. Canfora, \textit{Phys. Rev.} \textbf{D 88}, 065028
(2013).

\bibitem{canfora4} F. Canfora, F. Correa, J. Zanelli, \textit{Phys. Rev.}
\textbf{D 90}, 085002 (2014).

\bibitem{yang1} S. Chen, Y. Li, Y. Yang, \textit{Phys. Rev. }\textbf{D 89}
(2014), 025007.

\bibitem{canfora5} F. Canfora, M. Kurkov, M. Di Mauro, A. Naddeo, \textit{%
Eur.Phys.J.} \textbf{C75} (2015) 9, 443.

\bibitem{yang2} S. Chen, Y. Yang, \textit{Nucl. Phys.} \textbf{B 904} (2016)
470.

\bibitem{canfora6} E. Ayon-Beato, F. Canfora, J. Zanelli, \textit{Phys. Lett}%
. \textbf{B 752}, (2016) 201-205.

\bibitem{canfora7} F. Canfora, G. Tallarita, \textit{Phys. Rev}. \textbf{D}
\textbf{94} (2016), 025037.

\bibitem{canfora8} F. Canfora, G. Tallarita, \textit{Phys. Rev}. \textbf{D}
\textbf{91} (2015), 085033.

\bibitem{canfora9} F. Canfora, G. Tallarita, JHEP 1409 (2014) 136.

\bibitem{canfora10} F. Canfora, A. Paliathanasis,\ T. Taves, J. Zanelli,
\textit{Phys. Rev}. \textbf{D} \textbf{95} (2017), 065032.

\bibitem{canfora11} F. Canfora, G. Tallarita, \textit{Nucl. Phys.} \textbf{B
921} (2017) 394.

\bibitem{canfora12} P. Alvarez, F. Canfora, N. Dimakis, A. Paliathanasis,
arXiv:1707.07421, \textit{Phys. Lett}. \textbf{B 773} (2017) 401-407.


\bibitem{herdeiro} C. Herdeiro, V. Paturyan, E. Radu, D. H. Tchrakian,
\textit{Phys. Lett}. \textbf{B 772}, (2017), 63-69

\bibitem{AW} T. Andrade and B. Withers, JHEP05 (2014) 101

\bibitem{Gouteraux} B. Gouteraux, JHEP04 (2014) 181

\bibitem{CCPS} M. M. Caldarelli, A. Christodoulou, T. Papadimitriou and K.
Skenderis, JHEP1704 (2017) 001

\bibitem{gibbons2003} G.W.~Gibbons, \textit{Phys. Lett}. \textbf{B566}, 171
(2003)

\bibitem{Lemos95} J.~P.~S. Lemos, Class. Quantum Grav., 12 (1995) 1081

\bibitem{HuaLia95} C.-g. Huang and C.-b. Liang, Phys. Lett. \textrm{A}, 201
(1995) 27

\bibitem{Mann97} R.~B. Mann, Class. Quantum Grav. 14 (1997) L109

\bibitem{BLP} D. Brill, J. Louko and P. Peld\'an, Phys. Rev D56 (1997)
3600

\bibitem{Vanzo} L. Vanzo, Phys. Rev. D56 (1997) 6475

\bibitem{HawPag83}
S.~W. Hawking and D.~N. Page, Commun. Math. Phys. 87 (1083) 577

\bibitem{GibHaw77}
G.~W. Gibbons and S.~W. Hawking, Phys. Rev. {\rm D} 15 (1977) 2738

\bibitem{BalKra99}
V.~Balasubramanian and P.~Kraus, Commun. Math. Phys. 208 (1999),413

\bibitem{EmpJohMye99}
R.~Emparan, C.~V. Johnson, and R.~C. Myers, Phys. Rev. {\rm D} 60 (1999) 104001

\bibitem{Mann99}
R.~B. Mann, Phys. Rev. {\rm D}, 60 (1999) 104047

\bibitem{Gubser08}
S.~S. Gubser, Phys. Rev. Lett. 101 (2008) 191601

\bibitem{MarTroZan04}
C.~Mart\'{\i}nez, R.~Troncoso, and J.~Zanelli, Phys. Rev. {\rm D} 70 (2004) 084035


\bibitem{hernan} H.~A.~Gonzalez, M.~Hassaine and C.~Martinez, Phys.\ Rev.\
D80 (2009) 104008







\end{thebibliography}

\end{document}